\documentclass[aps,twocolumn,float,prl,psfig]{revtex4}

\usepackage{color}

\input epsf
\usepackage{graphicx}
\newcommand{\beq}{\begin{equation}}
\newcommand{\beqa}{\begin{eqnarray}}
		  \newcommand{\eeq}{\end{equation}}
\newcommand{\eeqa}{\end{eqnarray}}

\newcommand{\lsim}{\lesssim}
\newcommand{\gsim}{\gtrsim}

\newcommand{\lmk}{\left(}
\newcommand{\rmk}{\right)}

\newcommand{\lkk}{\left[}
\newcommand{\rkk}{\right]}
\newcommand{\lla}{\left\langle}
\newcommand{\p}{\partial}
\newcommand{\rra}{\right\rangle}
\newcommand{\so}{M_\odot}
\newcommand{\mch}{{\cal M}}

\newcommand{\tob}{T_{\rm o}}

\begin{document}
\title{Measuring the Galactic Binary Fluxes with LISA: \\  Metamorphoses and Disappearances of White Dwarf Binaries} 
%
%
%
\author{Naoki Seto }
\affiliation{Department of Physics, Kyoto University, 
Kyoto 606-8502, Japan
}
\date{\today}
%
%
%
%
%
%
\begin{abstract}
The space gravitational wave detector LISA is expected to detect $\sim10^4$ of nearly monochromatic binaries, after $\sim 10$\-yr operation.  We propose to measure the inspiral/outspiral binary fluxes  in the frequency space, by processing tiny frequency drifts of these numerous binaries.  Rich astrophysical information is encoded in  the frequency dependencies of the two fluxes, and  we can read the long-term evolution of white dwarf binaries, resulting in metamorphoses or disappearances.  This measurement will thus  help us to deepen our understanding on the strongly interacting exotic objects.  Using a  simplified model for the  frequency drift speeds, we discuss the primary aspects of the flux measurement,  including the prospects with LISA.

\end{abstract}

\maketitle

{\it Introduction.---}
Galactic ultra-compact binaries (orbital periods less than $\sim 10$\,min) are secure and  important observational targets for the space gravitational wave  (GW) interferometer LISA \cite{LISA:2017pwj,Cornish:2018dyw}.  They are also promising systems for multi-messenger observations \cite{Nelemans:2003ha,Korol:2018wep}.  By efficiently analyzing their data,  we will be able to obtain fruitful information on strongly interacting exotic objects.

Most of these ultra-compact binaries would be detached white dwarf binaries (WDBs) and AM CVn-type systems, both emitting nearly monochromatic GWs \cite{Hils:1990vc,Nelemans:2001nr,Nissanke:2012eh,Kremer:2017xrg}. The formers are  at the inspiral phase (${\dot f}>0$, $f$: GW frequency),
 and eventually their less massive white dwarfs fill the Roche-lobes, initiating the mass transfer.   In the basic picture, after this stage, the subsequent evolution bifurcates into two branches;  survival or disappearance \cite{p67,Nelemans:2001nr,s10,Nissanke:2012eh}.
If the mass transfer is stable, a WDB  morphs into an AM CVn-type system, and its frequency turns  into outspiral  (${\dot f}<0$), keeping the Roche-lobe overflow.   If the mass transfer is unstable, a WDB merges shortly, possibly accompanying an explosion event (e.g. type Ia supernova).  But, at present, our understanding on the bifurcation (e.g. branching ratio) is quite limited, due to the lack of observational knowledge and the theoretically  formidable physics on the strongly interacting compact objects \cite{Marsh:2003rd,s10}.

Even operating LISA for ten years, we are unlikely to observe a single  WDB merger in the Galaxy, given its estimated merger rate $O(10^{-2})$\,yr \cite{Nissanke:2012eh}.  However, after such an operation period, LISA will measure  small frequency drifts $\dot f$ for $\sim 10^4$ of the ultra-compact binaries  \cite{Nelemans:2003ha,Korol:2018wep}.

  In this letter, we propose to observationally determine the inspiral/outspiral binary fluxes, by using these swarm of binaries. 
 We point out the importance of the frequency dependencies of the two fluxes,  to statistically  follow the destinies of the WDBs.  
Below, combining the basic picture for  WDB evolution and a simplified model for the drift speed $\dot f$, we clarify  the primary aspects of the flux measurement at $f\gsim $5mHz.

 In fact, AM CVn systems are considered to be generated also from hybrid  binaries of white dwarfs and  nondegenerate helium stars.  Since they will emits GWs at most $\sim 3$mHz \cite{Nelemans:2001nr,Nissanke:2012eh}, we ignore this component below.

{\it Binary fluxes.---}
In the basic picture, 
by tracing flows of inspiral binaries  (see Fig. 1),  we can easily understand their continuity equation in the frequency space, at the large number limit  
\beq \label{b1}
{\p_t \rho_+}(f,t)+\p_f F_+(f,t)=\Sigma_{IJ}(f,t)-\Sigma_{M}(f,t)-\Sigma_{T}(f,t).
\eeq
Here  $\rho_+$ is the number density of inspiral binaries  and $F_+$ is their  flux.  The three non-negative  quantities $\Sigma_{IJ}$, $\Sigma_M$ and $\Sigma_T$ are the injection, merger and turnover rates (in units of $\rm Hz^{-1}s^{-1}$).
  In the basic picture, the outspiral flux $F_-(\le 0)$  is    sourced by the turnover rate (see Fig .1) and  described by 
$
{\p_t \rho_-}+\p_f F_-=\Sigma_{T}, 
$ {(ignoring potential disappearances after turnovers)}.

Our target band is $f\gsim 5$mHz and almost all the WDBs there are expected to be generated at lower frequencies \cite{Nissanke:2012eh}. We thus put $\Sigma_{IJ}=0$.  
Then,  from Eq. (\ref{b1}), we have
\begin{eqnarray}\label{b2}
F_+(f_1,t)-F_+(f_2,t)=\int_{f_1}^{f_2} \lkk  \Sigma_{T}(f,t)+\Sigma_{M}(f,t)\rkk  df\nonumber\\
+   \frac{\p}{\p t}\int_{f_1}^{f_2} \rho_{+}(f,t) df
\end{eqnarray}
The last terms is the correction caused by the time variation of the inspiral flux $F_+$.  Considering the Galaxy-wide binary formation and the delay time distribution before chirping up to $f\sim 5$mHz, the flux $F_+(f,t)$  (after suppressing the Poisson fluctuation) is expected to change slowly at the Hubble timescale $t_H\sim 10^{10}$yr \cite{Nelemans:2003ha,Lamberts:2019nyk}.  Then, in Eq. (\ref{b2}),  the last correction term will be $\sim t_d/t_H\sim 10^{-4}$ times smaller than the term $F_+(f_1,t)$.  Here  $t_d\sim 10^6$yr is the characteristic transition time from $f_1$ to $f_2$ (above $\sim 5$mHz).  As we see later,  the Poisson fluctuation of the fluxes (more than $\sqrt{10^{-4}}=10^{-2}$) will  completely mask  the correction term of this level. We thus drop the time dependence of variables,  and obtain 
\beq
F_+(f_1)-F_+(f_2)=\int_{f_1}^{f_2}\lkk  \Sigma_{T}(f)+\Sigma_{M}(f) \rkk df
\eeq
and similarly
$
F_-(f_1)-F_-(f_2)=-\int_{f_1}^{f_2} \Sigma_{T}(f) df .
$

By observationally measuring  the frequency dependencies of the two fluxes $F_\pm(f)$, we can separately estimate the two rates $\Sigma_M(f)$ and $\Sigma_T(f)$ at some frequency resolutions.  
This is the central part of the present proposal.  Note that the conservation equations have been sometimes  used theoretically,  mainly for estimating  the number densities of binaries (see e.g. \cite{Seto:2002dz,Farmer:2003pa,Nissanke:2012eh}).  But we  use these equations in a completely different way.

\begin{figure}
\vspace {.5cm} 
 \includegraphics[width=5.5cm,angle=270,clip]{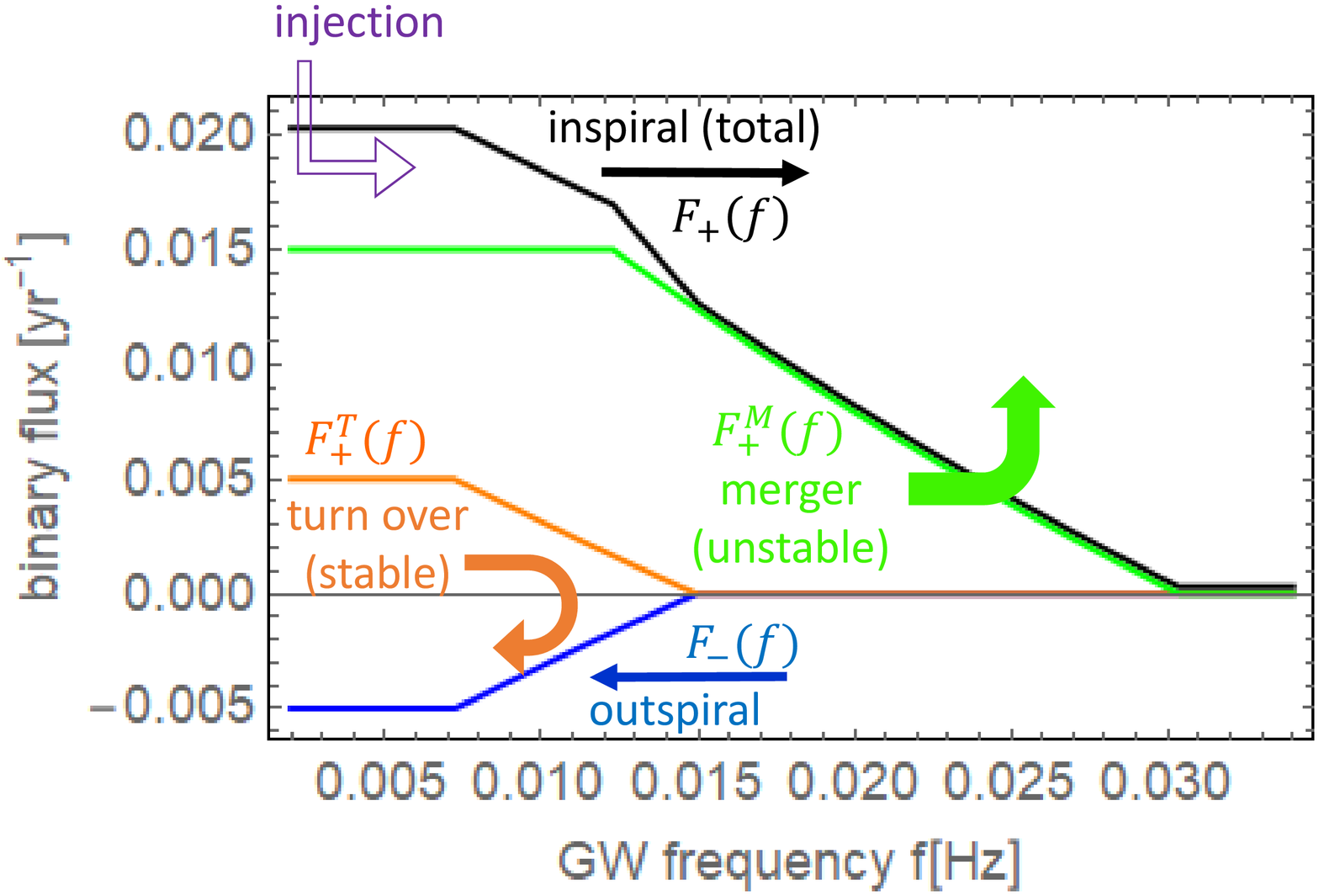}
 \caption{ The schematic picture  for the inspiral/outspiral  binary fluxes in the frequency space. The  black curve shows the total inspiral flux  $F_+(f)=F_+^M(f)+F_+^T(f)$  that is  subdivided into  the merger (green) and turnover (orange) components.  The latter generates the outspiral flux $F_-(f)=-F_+^T(f)$  (blue).  The information of the merger and turnover is clearly imprinted in  the frequency dependencies of  the two fluxes $F_\pm (f)$.
}
\end{figure}


{\it Evolution of individual binaries.---}
Next, to discuss the flux measurement more concretely,  we introduce a simplified model for the drift speed $\dot f$ \cite{p67,Nelemans:2003ha} (see also \cite{Gokhale:2006yn,Fuller:2014ika,McNeill:2019rct}).  This model would be  a workable approximation to the  steadily drifting binaries, except for the stages close to the merger or turnover frequencies. { In fact,  the binaries around the turn over ${\dot f}\sim0$ would show somewhat complicated time evolution \cite{Kaplan:2012hu,Tauris:2018kzq}.  But these binaries individually have small contributions ($\propto {\dot f}$) to the overall fluxes $F_\pm(f)$.}  We should also stress that, at actual observational measurement of the fluxes, we do not need  detailed theoretical models for the drift speeds. 

We first describe our simplified drift model for a circular inspiraling  WDB (${\dot f}>0$).
We denote its two initial masses  by $m_1$ and $m_2$ with $m_1\le m_2$, and define the initial mass ratio $q\equiv m_1/m_2\le 1$.   

Under the point particle approximation with the orbital separation $a$, the GW frequency $f$ 
 is given by $f=\pi^{-1}[G(m_1+m_2)a^{-3}]^{1/2}$ and orbital angular momentum  by $J=G^{1/2}a^{1/2}(m_1+m_2)^{-1/2}m_1 m_2$.  Due to the angular momentum loss by GW emission, we have 
\beq \label{df}
\frac{\dot f}{3f}=-\frac12\frac{\dot a}{a}= - \frac{J}{({\dot J})_{\rm gw}} =\frac{32 G^{5/3} \pi^{8/3} \mch^{5/3} f^{8/3}}{5c^5} \equiv t_{\rm gw}^{-1}  \label{tgw}
\eeq
with the chirp mass $\mch\equiv(m_1m_2)^{3/5}(m_1+m_2)^{-1/5}$.

\begin{figure}
\vspace {.5cm} 
  \includegraphics[width=7.5cm]{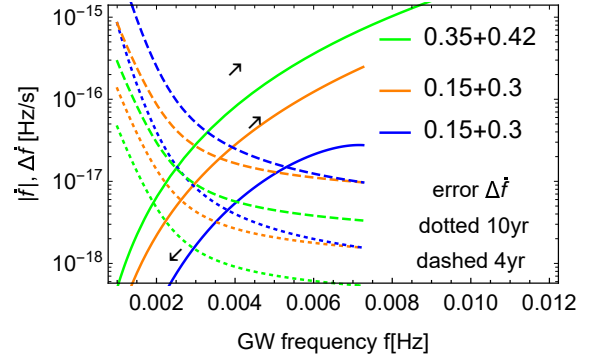}
 \caption{   The drift speeds $|\dot f|$ for the binaries with initial masses $0.35\so+0.46\so$ (green) and $0.15\so+0.3\so$ (orange: inspiral,  blue: outspiral phases).  The dotted and dashed curves show
 the estimations errors $\Delta{\dot f}$ for corresponding binaries  at $d=20$kpc with the observational periods  $\tob=10$yr and 4yr. }
\end{figure}


We use the mass-radius relation $r(m)$ for completely degenerate helium in \cite{vr}  originally given by P. Eggleton (see e.g. \cite{Deloye:2007uu} for thermal effects). 
The Roche lobe radius of the less massive one is roughly given by 
$
R_L\simeq {3^{-4/3}}\, 2\,a \,{m_1^{1/3}}{(m_1+m_2)^{-1/3}}
$ \cite{p67}.  
It shrinks, as the orbital separation $a$ decreases. Eventually the WD fills the Roche lobe at the separation with
$
R_L=r(m_1).
$
The GW frequency at this moment is given by
\beq
f_R(m_1)=\frac{2^{3/2}}{9\pi} \sqrt{\frac{G m_1}{r(m_1)^3}} \label{fr}
\eeq
as a function of the initial  mass $m_1$ \cite{Breivik:2017jip}.
Now the less massive WD becomes a donor of the mass transfer to the more massive WD.  If the initial mass ratio  satisfies the following inequality  ($\zeta(m)\equiv {d\ln r(m)}/{d\ln m}$)
\beq
q=\frac{m_1}{m_2}>\frac{3\zeta(m_{1})+5}6, \label{cf}
\eeq
the mass transfer is unstable and two WDs merge \cite{p67,Nelemans:2003ha}. Here we assumed the conservative mass transfer and efficient angular momentum redistribution to the orbital component.  The related physical parameters are not well understood at present  \cite{Marsh:2003rd,Gokhale:2006yn,Fuller:2014ika}, and our flux approach would provide us with useful information. 
For idealized cold Fermi gas at the non-relativistic limit, we have $\zeta=-1/3$ and $q>2/3$ for Eq. (\ref{cf}). 

In Fig. 2, with the green curve, we show the model prediction $\dot f$ for WDB with initial masses $0.35\so+0.42\so$ ($q=5/6$).  
This binary satisfies the unstable condition (\ref{cf})  and merges around $f_R(m_1)=17.8$mHz ($2.4\times 10^6$yr after passing 5mHz). 

If the condition (\ref{cf}) does not hold, the mass transfer is stable and the binary becomes an AM CVn-type system, turning from inspiral (${\dot f}>0$)  to outspiral  (${\dot f}<0$) in the frequency space.  

During the outspiral phase, the donor continuously fills the Roche lobe and its decreasing  mass is given by the GW frequency as $m_{1e}(f)=f^{-1}_R(f)$ with the inverse relation of Eq. (\ref{fr})  \cite{Breivik:2017jip}.
 For the accretor, we have 
$
m_{2e}(f)=m_1+m_2-m_{1e}(f).
$ 
Including the effects of the mass transfer, the frequency derivative of the outspiral phase  is given by  
\beq  \label{evm}
\frac{\dot f}{f}=-\frac32\frac{\dot a}{a}=\frac32  \lmk  \zeta(m_{1e})-\frac13\rmk  \lmk\frac{\zeta(m_{1e})}2+\frac56-\frac{m_{1e}}{m_{2e}}  \rmk t_{\rm gw}^{-1} 
\eeq
where the chirp mass in $t_{\rm gw}$ should be evaluated with the two evolved masses.

In Fig. 2, we show $\dot f$ for  a WDB  with initial masses $0.15\so+0.3\so$ ($q=1/2$ and $\zeta(m_1)=-0.33)$.  In the basic picture, this binary initially moves on the orange curve up to $f_R(m_1)=7.3$mHz. With stable mass transfer, it starts outspiral along the blue curve.   Due to the effects of the chirp mass and the second parenthesis  in Eq. (\ref{evm}),  the outspiral rate is much smaller than the inspiral rate.  It takes $1.1\times 10^6$yr  for this binary to move from 5.0mHz up to 7.3mHz,  and $5.9\times 10^6$yr to go back from 7.3mHz down to 5.0mHz.

In  reality, a relatively diffuse envelope of the donor could be stripped at the late inspiral phase \cite{Kaplan:2012hu,Tauris:2018kzq}.  But this would not change the concept of the flux approach (e.g. by using an appropriate relation $r(m)$ at the stage of interpreting the measured fluxes).

{\it Flux model.--- }
We now discuss the frequency dependence of the Galactic  inspiral and outspiral fluxes. For the former, we put the total value 
\beq
F_+(3{\rm mHz})=0.02{\rm yr^{-1}}
\eeq
at 3mHz with no additional injection above this frequency \cite{Nissanke:2012eh}. This flux  is divided into the merger and turn-over components   $F^M_+$ and $F_+^T$ (see Fig .1).

For the merger flux,  we set $F_+^M(3{\rm mHz})=0.015{\rm yr^{-1}}$, following \cite{Nissanke:2012eh} ($\sim 10^3$ larger than double neutron stars \cite{Seto:2019gtq}).  For its mass distribution, we fix the initial ratio  at $q_M=5/6$ and assume a flat profile for $m_1$ in the range    $0.25 \so \le m_1\le 0.54 \so$.  Here we set the massive end $0.54\so$ so that the characteristic frequency $f_R(0.54\so)=30$mHz is close to the highest WDB frequency predicted in \cite{Nissanke:2012eh}.  The lower end was chosen   somewhat arbitrarily with $f_R(0.25\so)=12$mHz.  Actually, the green curve in Fig. 1 shows the flux $F_+^M(f)$ obtained for the present setting. The nearly straight-line structure at 12-30mHz is due to the approximately  linear relation $f_R(m)$ in the relevant mass range.

 For the turnover flux, we assume $F_+^T(3{\rm mHz})=0.005{\rm yr^{-1}}$ as a model parameter  \cite{Nelemans:2001nr,Nissanke:2012eh}. For its mass distribution, we fix $q_T=1/2$ with a flat profile for $m_1$  in the range   $0.15 \so \le m_1\le 0.3 \so$.  As a precaution, we set the lower end to the very small value with $f_R(0.15\so)=7.3$mHz which corresponds to the minimum turnover/merger frequency $f_{l}$ in Fig. 1.  This frequency $f_l$ is important for the flux analysis and worth further study. {We also have $f_R(0.13\so)=6.3$mHz  and $f_R(0.17\so)=8.2$mHz for two different masses.}
The upper mass $0.3\so$ was selected to match the highest frequency of the Galactic AM CVns predicted in \cite{Nissanke:2012eh} with $f_R(0.3\so)=15$mHz.  
In Fig. 1, the orange curve shows the resultant turnover flux $F_+^M(f)$.  The outspiral flux (blue curve) is  given by $F_-(f)=-F_+^T(f)$ in the simplified picture.
{We should comment that  the magnitude of the turnover flux $F_+^T(3{\rm mHz})$ is  more uncertain than the merger flux  $F_+^M(3{\rm mHz})$. In a pessimistic model with an inefficient angular momentum redistribution  \cite{Nelemans:2001nr,Nissanke:2012eh}, the flux $F_+^T(3{\rm mHz})$ could be two order of magnitude smaller than the value adopted above.  } 

From the fluxes $F_\pm(f)$ and the drift speeds $\dot f$ for the composing binaries, we can evaluate the number densities of binaries per unit frequency interval $\rho_\pm(f)=F_+(f)/\overline{ {\dot f}_+ }(f)$ with the number weighted mean drift speed $\overline{ {\dot f}_+ }$.  In Fig. 3, we present our numerical results. In contrast to the fluxes in Fig. 1, the magnitudes of the orange and blue curves are not the same, reflecting the difference between the inspiral/outspiral speeds $\dot f$  as in Fig. 2.  At $f<f_l=7.3$mHz,  we have  the well-known form  $\rho_+(f)\propto f^{-11/3}$ simply determined  by Eq. (\ref{df}) \cite{Seto:2002dz,Nissanke:2012eh}.
Above 5mHz, the total numbers of the inspiral and outspiral binaries are estimated to be  6700 and 11800. 
{
If we decrease  the mass ratios $(q_M,q_T)$ by 10\% from the original setting $(5/6,1/2)$,  the larger components $m_2$  are increased  by 10\%.   With this modification, the profiles $F_\pm(f)$ in Fig. 1 are unchanged, depending basically on the distribution of $m_1$. But  the total numbers of binaries above 5mHz shrink to 6200 and 10400 respectively for inspiral and outspiral binaries,  because of  higher chirp rates $|{\dot f}|$.  
 }

\begin{figure}
 \includegraphics[width=6.5cm]{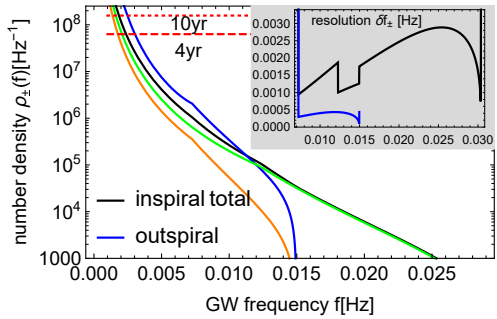}
 \caption{The number densities of binaries per unit frequency.   The black line is that  for  total inspiral binaries $\rho_+(f)$  composed by  the merger (green) and turnover (orange) components.  The blue curve is $\rho_-(f)$ for the outspiral binaries. The horizontal red lines  show the critical densities for the source confusion. The inset  shows the characteristic frequency resolutions $\delta f_\pm$  in Eq. (\ref{cres}).  The relative Poisson fluctuations are given by $[\rho_\pm(f)\delta f_\pm]^{-1/2}$.  }
\end{figure}


{\it GW observation and flux measurement.---}
We now discuss how to measure the two binary fluxes $F_\pm(f)$ with LISA. Our basic procedure will be  to firstly identify a large number of binaries by fitting their parameters including $\dot f$, and subsequently  calculate the fluxes using the detected binaries.

\begin{figure}[t]
 \includegraphics[width=6.5cm]{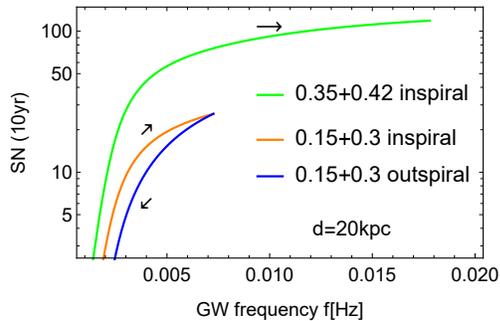}
 \caption{  The angular-averaged signal-to noise ratios for WDBs at $d=$20\,kpc and $\tob=10$yr.   The green curve is for the initial masses $0.35\so+0.42 \so$, resulting in the merger.  The WDB initially with $0.15\so+0.3 \so$   turns over  at  7.2\,mHz. }
\end{figure}
The angular-averaged strain amplitude of a nearly monochromatic binary at the distance $d$ is given by \cite{Cornish:2018dyw}
\beq
h=\frac{8 (G \mch)^{5/3}\pi^{2/3} f^{2/3}}{5^{1/2}c^{8/3}d}.
\eeq
Note that for an outspiral binary with $m_{1e}\ll m_{2e}$, we have
$\mch^{5/3}\sim m_{1e} m_{2e}^{2/3}\sim m_{1e}m_2^{2/3}$  and the amplitude depends  weakly on the assumption on the mass conservation.  The   angular-averaged signal-to-noise ratio is  estimated  to be 
\beq
SN={h\tob^{1/2}}/{S_n(f,\tob)^{1/2}}\label{amp}
\eeq
with the observational time $\tob$ and the strain noise $S_n(f,\tob)$ composed by the instrumental and confusion noises.  Here we included the $\tob$ dependence of the confusion noise \cite{Cornish:2018dyw,Seto:2019gtq}.  

In Fig. 4, we show the  signal-to-noise ratios  for binaries at $d=20$kpc and $\tob=10$yr.  Almost all  Galactic binaries have distances less than 20kpc \cite{Seto:2002dz,Nelemans:2003ha,Lamberts:2019nyk}. The blue curve (for the  initial masses $0.15\so+0.3\so$) can be regarded as the weakest signal emitter in the relevant frequencies regime (and smallest $|{\dot f}|$ except for those around the turnover).  Since an edge-on binary has $\sqrt{5}/4\sim 0.6$ times smaller amplitude than Eq. (\ref{amp}), we might miss some of outspiral binaries at $f\lsim 5$mHz with $SN\lsim 10$. 

For $\tob\gsim2$yr,  the measurement error for the drift speed is estimated to be 
\beq
\Delta{\dot f}=4 SN^{-1} \tob^{-2},
\eeq
 and depends strongly on $\tob
$ \cite{Takahashi:2002ky}.  In Fig. 2, with  the dotted and dashed curves, we show  the errors $\Delta {\dot f}$ for binaries identical to those in Fig. 4.  For $\tob\sim 10$yr, LISA is likely to have a resolution $\Delta {\dot f}/{\dot f}\lsim 0.2$ at $f\gsim 5$mHz, even for the smallest steady  speed $\dot f$ (blue curve). 


Here we briefly comment on the potential signal overlapping. For each drifting binary, the number of fitting parameters is eight, and we need two frequency bins to determine them (using four complex numbers from two data channels) \cite{Crowder:2006eu}.  Therefore, the densities should be  $
\rho_+(f)+\rho_-(f)\lsim \tob/2$ for resolving binaries.  As shown in Fig. 3,   for $\tob\gsim 4$yr, the signal confusion would not be a fundamental problem at $f \gsim 5$mHz (see also \cite{Lamberts:2019nyk}).

Next we discuss how to estimate   the inspiral flux $F_+(f)$ at a frequency  $f$.  {We can make almost the same argument for the outspiral flux  $F_+(f)$. }  Let us suppose that there are  altogether   $N_+^{\delta f}$  inspiral binaries in the  frequency range $[f-\delta f/2,f+\delta f/2]$ with $\delta f \ll f$.  With their label $i$,  the inspiral  flux can be estimated as 
\beq
F_{+}^{\delta f}(f)=\sum_{i =1}^{N_+^{\delta f}} \frac{{\dot f}_i}{\delta f}.
\eeq
{As we mention earlier, the binaries around turnover ${\dot f}\sim0$  individually have small contributions to this expression.}
We have the expectation values $\lla N_+^{\delta f} \rra\simeq \rho_+(f)\delta f$ and    $\lla F_{+}^{\delta f}(f)\rra\simeq F_+(f)=\rho_+(f) \overline{ {\dot f}_+ }(f)$ with the mean inspiral speed $\overline{ {\dot f}_+ }(f)$.  
The latter $\lla F_{+}^{\delta f}(f)\rra$ is independent of the width $\delta f$, but it has a statistical fluctuation $\Delta F_{+}^{\delta f}(f)$ in actual data reduction, due to the finiteness of the sample.   More specifically, we can write down
\beq
\frac{\Delta F_{+}^{\delta f}(f)}{F_+(f)}\sim (N_+^{\delta f})^{-1/2}  \lmk 1+ \frac{\sigma_{sc}}{ \overline{ {\dot f}_+}} + \frac{\sigma_{obs}}{ \overline{ {\dot f}_+ }}\rmk.
\eeq
The three terms in the last parenthesis originate from (i) the Poisson fluctuation of the sample number, (ii) the intrinsic scatter $\sigma_{sc}$ of the speed $\dot f$  and (iii) the  typical magnitude $\sigma_{obs}$ of the measurement error $\Delta {\dot f}$.  For $f\gsim 5$mHz and $\tob\sim 10$yr, the third one would be negligible (see Fig. 2). Assuming  $\sigma_{sc}\sim \overline{ {\dot f}_+}$, we have ${\Delta F_{+}^{\delta f}(f)}/{F_+(f)}\sim (N_+^{\delta f})^{-1/2}$ corresponding to the Poisson fluctuation.  {For example, with our model parameters, we have the inspiral and outspiral binaries of $N_+=2050$ and $N_-=3840$ in the range [5.5mHz,\,6.5mHz].  We thus measure the fluxes $F_\pm ({\rm 6mHz})$ with Poisson fluctuations  less than $\sim $3\%. Without injections,  mergers and turnovers in [3mHz,6.5mHz], we will have   $F_\pm ({\rm 3mHz})=F_\pm ({\rm 6mHz})$. }

 {As discussed earlier around Eqs. (2) and (3), at $f>f_l$, we also  want to finely resolve the frequency dependence of the flux $F_+^{\delta f}$ by taking a small width $\delta f$.}  But, at the same time, the statistical error should be suppressed. We can take a balance by choosing the width as 
$
\Delta F_+^{\delta f} \simeq | F_+(f+\delta f/2)-F_+(f-\delta f/2)|.
$
Similarly considering the outspiral flux, we have the approximate solutions as
\beq
\delta f_\pm=  \rho_\pm^{-1/3} | F_\pm|^{2/3} \left|{dF_\pm}/{df} \right|^{-2/3}.\label{cres}
\eeq
 As shown in Fig. 3, in the range  $dF_\pm /df\ne 0$, we have $\delta f_+ \sim $1-3mHz  and $\delta f_-\sim$ 0.2-0.4mHz.  {For  deriving Eq. (\ref{cres}),   we approximately  put $F_+(f+\delta f/2)-F_+(f-\delta f/2)\sim \delta f \cdot dF_+/df$.  In Fig. 3, this derivative  introduces the sharp features, reflecting the discontinuities of $dF_\pm(f)/df$ as seen in Fig. 1.   }

For these solutions $\delta f_\pm$, we have the corresponding numbers of binaries  $\rho_{\pm}(f)\delta f_{\pm}\sim$1000-100 in the 7-15mHz range and smaller at higher frequencies.  Therefore, the typical magnitude of the Poisson fluctuation is $\sim 100^{-1/2}\sim 0.1$. 

{\it Discussion.---}
In this letter, we proposed to measure the Galactic binary fluxes, by using $\sim 10^4$ of WDBs detected by LISA.  By studying the frequency dependencies (closely related to initial mass $m_1$) of the  fluxes at $f\ge f_l$, we can clearly follow how WDBs disappear or survive, affected by  physical processes on strongly interacting exotic objects.  To examine further details of the binary evolution beyond the basic picture, we could additionally use the distribution of the drift speeds $\dot f$. 

While untouched so far, the fluxes  at $ f<f_l$ would be also useful. We can check the stationary of the fluxes and study potential binary injections and disruptions  there.  To make a complete Galactic sample at relatively low frequency regime (e.g. $f\lsim 5$mHz),  other space interferometers (e.g. Taiji \cite{taiji} and TianQin \cite{Luo:2015ght}) could make important contributions, given the expected performance of LISA shown in Figs. 2 and 4.  {We can also employ Galactic structure  models to correct the contributions of distant binaries that have too small rates $|{\dot f}|$ or even too small amplitudes $h$  \cite{Korol:2018wep,Lamberts:2019nyk}}.

\begin{acknowledgments}
 This work is supported by JSPS Kakenhi Grant-in-Aid for Scientific Research
 (Nos. 17H06358 and 19K03870).
\end{acknowledgments}

\end{document}